\documentclass[aps,prb,10pt,twocolumn,amsmath,amssymb,citeautoscript,
               superscriptaddress, longbibliography]{revtex4-1}

\usepackage{graphicx}
\usepackage{epstopdf}
\usepackage{dcolumn}
\usepackage{array}
\usepackage{amssymb}
\usepackage{amsmath}
\usepackage{multirow}
\usepackage{braket}
\renewcommand{\vec}[1]{\mathbf{#1}}
\usepackage{bm}
\usepackage[hang]{footmisc}

\begin{document}
\title{Berry Flux Diagonalization: Application to Electric Polarization}
\author{John Bonini, David Vanderbilt, and Karin M. Rabe}
\affiliation{Department of Physics and Astronomy, Rutgers University, Piscataway, NJ 08854-8019}
\date{\today}
\begin{abstract}
  The switching polarization of a ferroelectric is determined by the
  current that flows as the system is switched
  between two variants. Computation of the switching polarization in crystal
  systems has been enabled by the modern theory of polarization, where it is
  expressed in terms of a change in Berry phase from the initial state to the final state. It is
  straightforward to compute this change of phase modulo $2\pi$,
  then requiring a branch choice to specify the predicted switching polarization.
  The measured switching polarization depends on the actual path along which the
  the system is switched, which in general involves nucleation and growth of domains
  and is therefore quite complex.
  In this work we present a first principles
    approach for predicting the switching polarization
    that requires only a knowledge of the initial and final states, based on the
    empirical observation that for most ferroelectrics, the observed
    polarization change is the same as for a path involving minimal evolution of
    the state.
To compute the change along a generic minimal path, we decompose the change of Berry
phase into many small contributions, each
much less than $2\pi$, allowing for a natural resolution of the
branch choice.
We show that for typical ferroelectrics, including those
  that would have otherwise required a densely sampled path, this technique
  allows the switching polarization to be computed without any need for
  intermediate sampling between oppositely polarized states.
\end{abstract}

\maketitle

\section{Introduction}
Bistable systems with a change in electric polarization on switching between the
two states are of central importance in functional material and device design.
The most familiar of such systems are ferroelectrics, with two or more
symmetry-related polar insulating states\cite{Lines_2001}. Switching in systems
in which the two states are not symmetry related, for example in
antiferroelectrics or heterostructures, is also of great interest for novel
devices\cite{doi:10.1002/9783527654864.ch7, PhysRevLett.84.5636}.

First principles prediction of the switching polarization in periodic systems is
based on the modern theory of polarization, which expresses the polarization
change between two states in terms of the change in Berry phase as the system
evolves along a specified adiabatic path.
\cite{king-smith93_theor_polar_cryst_solid, vanderbilt_2018_3} From knowledge of
the initial and final states, the polarization change is determined modulo the
``quantum of polarization'' ($e \vec{R}/\Omega$), where $e$ is the charge of an
electron, $\vec{R}$ is a lattice vector, and $\Omega$ is the volume of the unit
cell. Additional information about the path would be needed to determined which
of the allowed values corresponds to any given path.

Since the path for a process such as electric field switching of a ferroelectric
generally involves nucleation and growth of domains, beyond the scope of current
first-principles computation, it might at first seem that first-principles
prediction of the switching polarization should not be possible. However, it is
an empirical fact that good agreement with experimental observation has been
obtained for many ferroelectrics by computing the polarization change along a
fictitious minimal path\cite{PhysRevB.96.035143,
  PhysRevB.71.014113,spaldin12_begin_guide_to_moder_theor_polar}, usually
constructed by simple linear interpolation of the atomic positions of the up-
and down-polarized states, maintaining their lattice translational symmetries.
The polarization change along this fictitious path is then computed by sampling
densely enough along the path so that the polarization change for every step
along the path can be chosen (and is chosen) to be small compared to the quantum
of polarization. However, this method can be computationally intensive,
depending on the sampling density required. Moreover, for some systems, it might
be that not all the states on the simple linear interpolation path are
insulating, and additional effort is required to find an insulating adiabatic
path connecting the up and down states. As a result, this approach has proven to
be problematic for automated high-throughput
applications\cite{smidt20_autom_curat_first_princ_datab_ferroel}.

In this paper, we present a new method for predicting switching polarization
given only the initial and final states. Our approach uses information, computed from
the two sets of
ground state wavefunctions, that goes beyond that used in a conventional Berry
phase calculation. The key idea is to incorporate certain assumptions about the
physical path, eliminating the need to construct a fictitious path and perform
calculations for intermediate states.
We begin by discussing the method for the simplest case of the electronic
contribution to the switching polarization for a one-dimensional polar
insulator. We then generalize to three-dimensional materials and discuss the
ionic contribution to the polarization. Finally, first-principles results are
presented for a realistic benchmark system to illustrate the various aspects of
the method and to compare with the fictitious path method.
The approach presented here is not limited to computation of switching
polarization in ferroelectrics, but can be applied to the change in polarization
between two symmetry-inequivalent states, for example in antiferroelectrics,
heterostructures and pyroelectrics, and in the computation of the nonlinear
response of insulators to electric fields.

\section{Formalism}

\subsection{Background and notation}

We start by considering a one-dimensional crystal switching from initial state
$A$ to final state $B$ along a specified path, parameterized by $\lambda$, along
which the system remains insulating. According to the modern theory of
polarization,\cite{king-smith93_theor_polar_cryst_solid, RevModPhys.66.899,
  vanderbilt_2018_3, spaldin12_begin_guide_to_moder_theor_polar} the electronic
contribution to the change in polarization can be expressed as
\begin{equation}
\Delta P_{A\rightarrow B} = \frac{-e}{2\pi}\,\Phi
\label{eqn:DP}
\end{equation}
where $\Phi$ is the Berry flux
\begin{equation}
  \Phi = \int\int_{S} \Omega(k, \lambda) d\lambda dk
  \label{eqn:berryCurve}
\end{equation}
obtained by integrating the Berry curvature $\Omega(k, \lambda)$ over the region
$S$ with $\lambda_A \leq \lambda \leq \lambda_B$ and $-\pi/a < k \leq \pi/a$
(the first Brillouin zone).
Here the Berry curvature
\begin{equation}
  \Omega(k, \lambda) = \sum_{n}-2\mathrm{Im} \braket{\partial_{\lambda}u_{n}(k, \lambda)|\partial_{k}u_{n}(k, \lambda)}
\label{eqn:bcurv}
\end{equation}
is written in terms of the cell-periodic parts of the occupied Bloch
wavefunctions $\ket{u_{n}(k, \lambda)}$ and has been traced over the occupied
bands $n$. The existence of the derivatives in Eq.~(\ref{eqn:bcurv}) requires
choice of a ``smooth gauge'': that is, for a single occupied band, the $k$- and
$\lambda$-dependent phase of the wave functions $\ket{u(k, \lambda)}$ must be
chosen so that $\ket{u(k, \lambda)}$ is differentiable as a function of $k$ and
$\lambda$ over all of $S$. For multiple occupied bands, specification of a gauge
may involve a $(k,\lambda)$-dependent unitary rotation of the occupied bands.
Since physical observables like the change in polarization along a specified
path do not depend on the choice of gauge, we are free to choose a gauge for
which the $\ket{u_{n}(k,\lambda)}$ are periodic in $k$.

Application of Stoke's theorem gives
\begin{equation}
  \Phi = \oint_{C} \mathbf{A}(\mathbf{q}) \cdot d\mathbf{q}
  \label{eqn:berryPotential}
\end{equation}
where $C$ is the boundary of the surface $S$, $\mathbf{q}=(k, \lambda)$, 
and ${\bf A}(\vec{q})=(A_k,A_\lambda)$ is the
Berry potential given by
\begin{align}
    A_{k} &= \sum_{n}i\braket{u_{n}(k, \lambda)|\partial_{k}u_{n}(k, \lambda)}\,,
\label{eqn:bpot}\\
    A_{\lambda} &=  \sum_{n}i\braket{u_{n}(k, \lambda)|\partial_{\lambda}u_{n}(k, \lambda)}\,.
\end{align}
Since we have chosen a gauge 
periodic in $k$, 
we have $\ket{u_{n}(k+2\pi,\lambda)}=\ket{u_{n}(k,\lambda)}$.
Then, we have $A_\lambda (k+2\pi,\lambda)=A_\lambda(k,\lambda)$
and the contributions $\int_{0}^{1} A_{\lambda}(k, \lambda) d\lambda$ and
$\int_{1}^{0} A_{\lambda}(k+2\pi, \lambda) d\lambda$ from the two portions of
the path $C$ along the $\lambda$ direction cancel. The two remaining segments
take the form
\begin{equation}
   \phi_{\lambda} =
\int_{-\pi/a}^{\pi/a} A_{k}(k, \lambda) dk
  \label{eqn:phi_multi}
\end{equation}
and it follows that
\begin{equation}
\Phi = \phi_{\lambda_B} - \phi_{\lambda_A} \,.
\label{equ:Phi-diff}
\end{equation}
The electronic contribution to the change in 
polarization is then given by Eq.~(\ref{eqn:DP}).

While the evaluation of \ref{equ:Phi-diff} requires only wavefunctions on the
boundary of $S$, the equivalence of Eqs.~(\ref{eqn:berryCurve}) and
(\ref{equ:Phi-diff}) requires the existence of a smooth gauge on all of $S$ that
matches the choice of gauge on the boundary. If the gauge is required to be
smooth only on the boundary of $S$, without this additional constraint, then a
gauge transformation can change quantities such as
$\phi_{\lambda_ A}$ and $\phi_{\lambda_B}$ by multiples of $2\pi$.
\cite{spaldin12_begin_guide_to_moder_theor_polar,
  king-smith93_theor_polar_cryst_solid, vanderbilt_2018_3}
We refer to such quantities as ``gauge invariant modulo a quantum," in
distinction to quantities which are ``fully gauge invariant'', and to ``fully
gauge dependent'' quantities that can take on any value in a continuous range
with a change in gauge. In this case $\Phi$ in Eq.~(\ref{equ:Phi-diff}) is
determined only modulo $2\pi$, and the change in polarization is determined only
modulo the quantum of polarization $e\vec R /\Omega$.

In other words, Eq.~(\ref{eqn:berryCurve}) is the fundamental expression for the
change in polarization along a specified path. It depends on the wavefunctions
at all intermediate $k$ and $\lambda$ and is fully invariant under gauge
transformation of the wavefunctions. On the other hand, the Berry phase
difference Eq.~(\ref{equ:Phi-diff}) depends only on the wavefunctions on two
edges of the boundary of S, and under gauge transformation of these
wavefunctions is only gauge invariant modulo a quantum. It is equal to the
change in polarization along the specified path only if the gauge chosen on the
boundary is one that can be smoothly continued into the interior onto the
wavefunctions at all intermediate $k$ and $\lambda$.

The fictitious minimal path method, the most widely used method for resolving
the branch choice for the difference $\phi_{\lambda_B} - \phi_{\lambda_A}$ to
compute the change in polarization along a specified path, relies on sampling a
minimal path, usually obtained by linear interpolation, at intermediate values
of $\lambda$. The density of sampling increased until each new
$\phi_{\lambda_{j+1}}$ can be chosen such that
$|\phi_{\lambda_{j+1}}-\phi_{\lambda_j}|<<\pi$. With a sufficiently dense
sampling the branch choice identified by this procedure will match that of the
continuum formulation, giving the correct polarization change for this path. In
practice the computation of $\phi_{\lambda}$ requires a discretization in $k$.
Sec.~\ref{sec:discrete} provides more details on how the relevant quantities are
computed when states are sampled on a discrete mesh.

Here, we present an alternative approach to resolving the branch choice that
makes full use of the information contained in the initial and final states,
while eliminating the need for sampling at intermediate values of $\lambda$.
Moreover, this approach requires a $k$-space sampling no denser than that
required for the computation of the formal polarization. Like the previous
methods, the fully gauge-invariant quantity $\Phi$ is separated in to smaller
contributions that, while gauge-invariant only modulo $2\pi$ in principle, can
always be taken much smaller than $2\pi$ in practice. However, here the
construction relies only on the wavefunctions of the initial and final states,
without reference to intermediate steps along the path. The additional
assumption required for this procedure is that the initial and final states at
$\lambda_A$ and $\lambda_B$ must be similar enough that their gauges can be
aligned, in a sense to be described shortly. Essentially, the gauge alignment
procedure implements a minimal evolution of the electronic structure, in analogy
with the previously assumed minimal evolution of the ionic structure.

\subsection{\label{sec:gaugeclass} Gauge class}

We first consider the case of a single occupied band in 1D with Bloch states
$\ket{u(k)}$. Following Eq.~(\ref{equ:Phi-diff}), the Berry phase around the
Brillouin zone at a given $\lambda$ is given by
\begin{equation}
    \phi   =  \int_{-\pi/a}^{\pi/a}\braket{u(k)|i\partial_{k}u(k)}dk
  \label{eqn:phi_single}
\end{equation}
and, following the terminology introduced in the previous section, is gauge
invariant modulo a quantum: specification of a gauge that is smooth on the first
Brillouin zone and periodic in $k$ allows transformations of the form
$e^{-i\beta(k)}\ket{u(k)}$, where $\beta(k)$ is differentiable and
$\beta(k+2\pi/a)=\beta(k)+2 \pi n$ for some integer $n$, which changes $\phi$ by
$2\pi n$. For a given physical system, we can test whether two choices of gauge
$a$ and $b$ will produce the same value of $\phi$ by computing
\begin{equation}
\label{eqn:gamdef}
\gamma^{\mathrm{ab}}(k) = \braket{u^a(k)|u^b(k)} \,.
\end{equation}
Note that $\gamma^{\mathrm{ab}}(k)$ has exactly unit norm and is just
$e^{-i\beta(k)}$, where $\beta(k)$
describes the gauge change relating $a$ to $b$.
If $\gamma^{ab}(k)$
is smooth and its phase
does not wind by a nonzero
integer multiple of $2\pi$ as $k$ traverses the 1D Brillouin zone,
the two gauges will produce the same $\phi$, and can be said to
belong to the same ``gauge class.''

Next, we consider two crystals $A$ and $B$ with single occupied bands, each with
a smooth gauge, and ask whether their respective gauges belong to the
same gauge class in a similar sense.
With this motivation, we define, in analogy with Eq.~(\ref{eqn:gamdef}),
\begin{equation}
  \gamma^{AB}(k) = \braket{u^{A}(k)|u^{B}(k)}
  \label{equ:gamma}
\end{equation}
where $\gamma^{AB}(k)$ will generally not have unit norm. In fact, for this
procedure to be meaningful, systems $A$
and $B$ must be sufficiently closely related that the norm of
$\gamma^{AB}(k)$ remains nonzero everywhere in the Brillouin
zone. If the phase of
this $\gamma^{AB}(k)$ does not wind by a nonzero integer
multiple of $2\pi$, we consider their gauges to belong to the
same gauge class.

We are now in a position to introduce our key idea for the prediction of
the switching polarization from system $A$ to $B$.
We recall the
empirical fact, discussed in the Introduction, that good agreement with experimental observation has been
obtained for many ferroelectrics by computing the polarization change along a
fictitious minimal path. 
Our insight is that in general, along such paths, the wavefunction phases will evolve in
a minimal way that preserves the gauge class, so that the switching polarization
corresponds to the polarization difference of Eq.~(\ref{eqn:DP}) and
Eq.~(\ref{equ:Phi-diff}) with Berry phases $\phi^A$ and $\phi^B$ computed with the
requirement that the two gauges belong to the same gauge class. Crucially, the
branch-choice ambiguity in the individual $\phi^A$ and $\phi^B$ is no longer
present after the difference is taken.

The generalization to the multiband case is straightforward.
We define
\begin{equation}
  \gamma^{AB}(k) = \det M^{AB}(k)
\label{eq:gammul}
\end{equation}
where $M^{AB}(k)$ is the overlap matrix given by
\begin{equation}
  M^{AB}_{mn}(k) = \braket{u_{m}^A(k)|u_{n}^B(k)}
\end{equation}
for occupied band indices $m$ and $n$.
The gauges are said to belong to the same class if the
phase winding of $\gamma^{AB}(k)$ is zero.

One way to insure that gauges $A$ and $B$ belong to the same gauge class
is to align one to the other.  In the single-band case, the gauge of $B$ is
aligned to that of $A$ by taking $\chi(k)={\rm Im\,ln\,}\gamma^{AB}(k)$,
and then letting
\begin{equation}
|\tilde{u}^{B}(k)\rangle=e^{-i\chi (k)}\,|u^B(k)\rangle \,.
\label{eqn:align-one}
\end{equation}
As a result, the new $\tilde{\gamma}^{AB}(k)$ is real and positive, so
that there is clearly no winding. Similarly,
the multiband gauge alignment can be accomplished by carrying out
the singular value decomposition of $M^{AB}$ in Eq.~(\ref{eq:gammul}) as
$M^{AB}=V^\dagger \Sigma W$, where
$V$ and $W$ are unitary and $\Sigma$ is positive real diagonal.
Then the multiband analog of
$e^{i\chi}$ is $U=V^\dagger W$, and the gauge of $B$ is aligned
to that of $A$ by the transformation
\begin{equation}
|\tilde{u}_{n}^B\rangle=\sum_m (U^\dagger)_{mn}|u_{m}^B\rangle \,.
\label{eqn:align-multi}
\end{equation}
The new overlap matrix is then $\tilde{M}^{AB}=V^\dagger\Sigma V$,
whose determinant $\tilde{\gamma}^{AB}$ in Eq.~(\ref{eq:gammul}) is clearly
real and positive, thus eliminating the relative winding of gauge $B$ with
respect to $A$.

The physical interpretation of enforcing both systems to belong to
the same gauge class is that we are assuming a minimal evolution of
the electronic states.  If the initial and final states represent
exactly the same bulk system, the result of this procedure is
obviously that the change in polarization is zero. In this situation,
other choices of gauge that give a nonzero value describe a
physical path where one or more quanta of charge are pumped by a
lattice vector over an adiabatic cycle with the periodic system returning
to its initial state. In making the same-gauge-class assumption for systems
where the initial and final states are different, but still closely related, we
similarly identify a result which involves a minimal evolution of the state.
An analogy can be made with the implicit assumptions already being made when one
constructs a switching path for the ions. When comparing initial and final
states, one typically specifies which ion maps to which by minimizing displacements
between ions of the same species, e.g., such that no ion moves by more than
half a unit cell. The gauge alignment procedure described above does
something similar, mapping which
band goes with which by maximizing wavefunction
overlaps and eliminating phase differences at corresponding $k$-points.

\subsection{\label{sec:discrete} Discrete $k$ space}
In any numerical calculation, functions of $k$ must be
sampled on a discrete mesh in $k$.
In this case, we can again align the
gauge of $B$ to that of $A$ using Eq.~(\ref{eqn:align-one}) or
Eq.~(\ref{eqn:align-multi}), and compute the polarization difference
via Eq.~(\ref{equ:Phi-diff}).  However, in the discrete case
there is a new potential source of ambiguity coming from the
need to enforce smoothness with respect to $k$.
After discretization Eq.~(\ref{eqn:phi_multi}) becomes
\begin{equation}
\phi_{\lambda} = \mathrm{Im}\,\mathrm{ln}\,\mathrm{det}\prod_{i}
   M^{\lambda}(k_i,k_{i+1}) \,
  \label{eqn:discrete}
\end{equation}
where $M$ is the overlap matrix
\begin{equation}
  M_{mn}^{\lambda}(k_i,k_{i+1}) = \braket{u_{m}^{\lambda}(k_i)|u_{n}^{\lambda}(k_{i+1})} \,.
  \label{eqn:overlap}
\end{equation}
This $\phi_{\lambda}$ is gauge-invariant, but only up to an integer multiple of
$2\pi$. This is reflected by the $\mathrm{Im}\mathrm{ln}$ operation in
Eq.~(\ref{eqn:discrete}),
which will only result in a phase in the interval $-\pi < \phi_{\lambda} < \pi$.
If one is interested in this phase on its own (i.e., for
computing formal polarization) this makes perfect sense, since it is truly a
lattice valued quantity. However, our present goal is to compute the difference
in phase between two systems with the requirement that both systems are in the
same gauge class. For this purpose it is useful to rewrite
Eq.~(\ref{eqn:discrete}) in a form where values outside this interval are
possible (with the branch being determined by the gauge). To this end we rewrite
Eq.~(\ref{eqn:discrete}) as
\begin{equation}
\phi_{\lambda} = \sum_i {\cal A}_i(\lambda)
\label{equ:sum-of-A}
\end{equation}
where
\begin{equation}
{\cal A}_i(\lambda)= \mathrm{Im}\,\mathrm{ln}\,\mathrm{det}\,
M_\lambda(k_i,k_{i+1}) \,
\label{equ:discrete-A}
\end{equation}
is a discrete analog of the Berry connection $A_k$.
We choose a sufficiently fine $k$ mesh and a sufficiently smooth gauge so that
each ${\cal A}_{i}$ is much less than $\pi$ in magnitude; then
$\phi^A$ can be unambiguously
computed (for the chosen gauge).
We then choose the gauge in $B$ to be aligned to that of $A$. Assuming this also
results in a smooth gauge in $B$,
we could then confidently compute $\Delta P$ from
Eqs.~(\ref{eqn:DP}) and Eq.~(\ref{equ:Phi-diff}).

\subsection{\label{subsec:gauge_relax} Gauge invariant formulation}
The procedure described in the last section involved constructing a smooth gauge in
$A$, aligning the gauge in $B$, and then computing each $\phi_{\lambda}$ via
Eq.~(\ref{equ:sum-of-A}). This represents a straightforward, but also inconvenient,
means of applying the same gauge class assumption to a realistic calculation. In
this section and the next we will develop an equivalent procedure
that is more computationally efficient and does not require explicit
construction of smooth or aligned gauges.

First, we note that the value
obtained above is equivalent to evaluating
$\Phi$ as
\begin{equation}
\Phi=\sum_i \Delta{\cal A}_i
\label{equ:Phi-DA}
\end{equation}
where
\begin{equation}
\Delta{\cal A}_i={\cal A}_i(\lambda_B)-{\cal A}_i(\lambda_A)
\label{equ:DA}
\end{equation}
is the difference between Eq.~(\ref{equ:discrete-A}) evaluated at the initial
and final configurations (with the previously discussed gauge choices).
At present, it is required that $k$ has been sampled densely enough such
that each $\Delta{\cal A}_{i}$
is smaller in magnitude than $\pi$.

\begin{figure}
  \includegraphics[width=\linewidth]{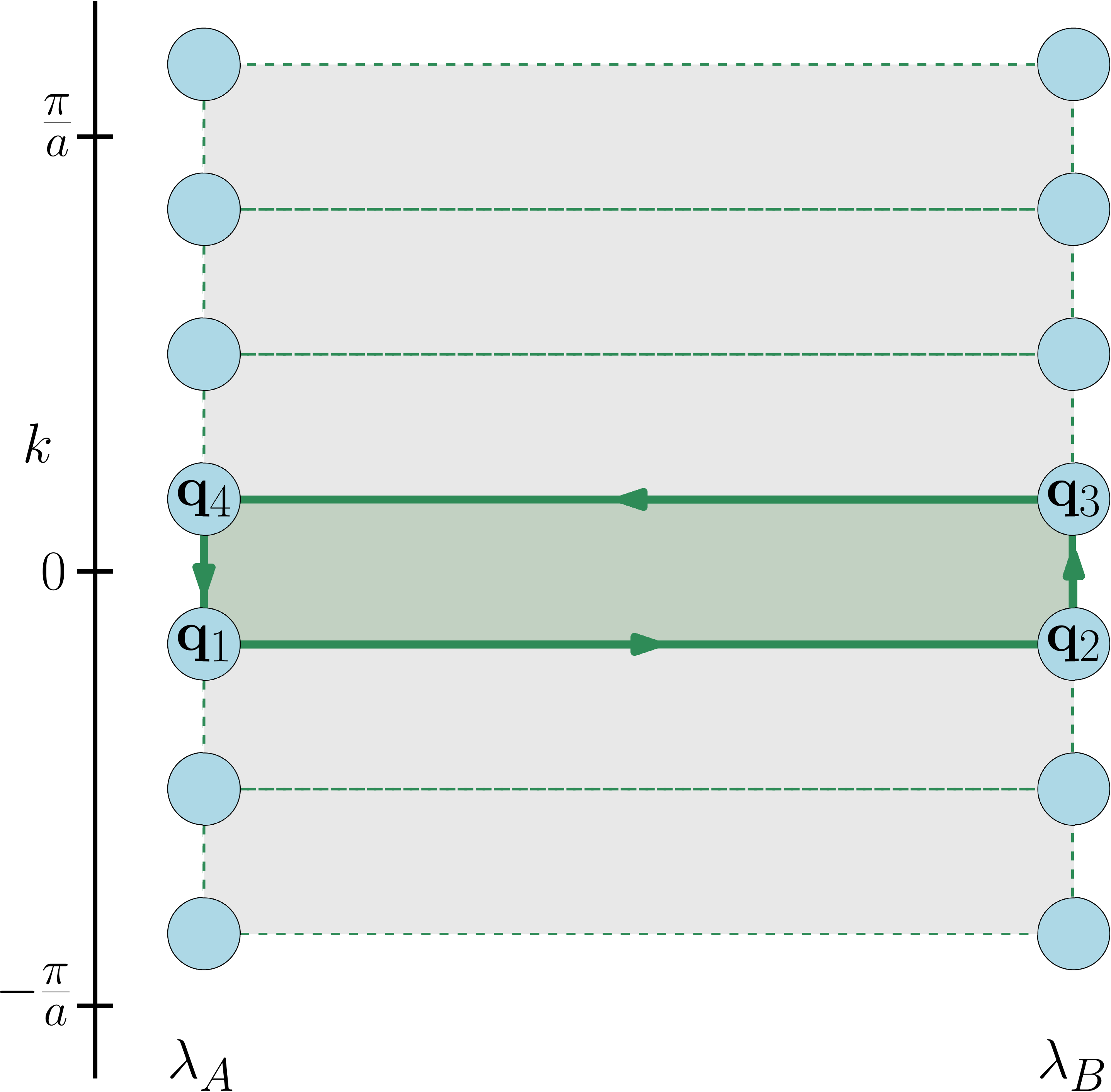}
  \caption{Sketch of the joint $(k, \lambda)$ space for computing a change in
    polarization between $\lambda_{A}$ and $\lambda_{B}$. Blue circles represent
    points where Bloch wavefunctions have been computed. The light grey box
    represents the surface $S$ that is integrated over in Eq.~(\ref{eqn:berryCurve}).
    Dotted green lines represent the plaquettes $i$ and the solid green lines
    represent the path on which the parallel transport procedure is performed
    around the green plaquette it encloses to obtain its contribution to
$P^{B} - P^{A}$.}
  \label{fig:k_lambda_space}
\end{figure}

We next note that the quantity $\Delta{\cal A}_{i}$
is equal to the discrete
Berry phase computed around the perimeter of the rectangular plaquette marked by
the green arrows in Fig.~(\ref{fig:k_lambda_space}). To see this, we denote the four corners
of this plaquette as
$\bm{q}_{1}=(k_i,\lambda_A)$, $\bm{q}_{2}=(k_i,\lambda_B)$,
$\bm{q}_{3}=(k_{i+1},\lambda_B)$, and $\bm{q}_{4}=(k_{i+1},\lambda_A)$, and refer
to it henceforth as plaquette $p$ located at $k_i=k_p$.
Defining the overlap matrices
\begin{equation}
M^{\braket{ij}}_{mn}=\braket{u_m(\bm{q}_i)|u_n(\bm{q}_j)} \,,
\label{equ:Mq}
\end{equation}
the four-point Berry phase about the loop, traced over occupied
bands, is
\begin{equation}
\phi^{p} = \mathrm{Im}\,\mathrm{ln}\,\mathrm{det}\,
[ M^{\braket{12}} M^{\braket{23}} M^{\braket{34}} M^{\braket{41}} ]\,.
\label{equ:pjk}
\end{equation}
This four-point Berry phase is equal to the Berry flux through
the plaquette, by the same Stoke's theorem argument used to relate Eq.~(\ref{eqn:berryCurve})
and Eq.~(\ref{eqn:berryPotential}).
This plaquette Berry flux, $\phi_{i}$, can be seen to be equal to $\Delta A_{i}$
computed with the gauges specified above
since the alignment of gauges insures that $M^{\braket{12}}$ and
$M^{\braket{34}}$ have real positive determinants, and thus don't contribute to
the phase being extracted by the $\mathrm{Im}\mathrm{ln}$ operation.
The advantage of
computing $\phi^{p}$ as in Eq.(~\ref{equ:pjk}) is
that
it is completely insensitive to the gauges used to represent the states at any
of the four $\vec{q}_{i}$ \cite{fukui_bcurve}. Using Eq.~(\ref{equ:Phi-DA}) we can write
$\Phi$ as the sum over plaquette Berry fluxes,
\begin{equation}
\Phi=\sum_{p} \phi^{p} \,.
\label{equ:Phi-phi}
\end{equation}

As the $\mathrm{Im}\mathrm{ln}$ operation suggests, $\phi^{p}$ is only gauge
invariant up to an integer multiple of $2\pi$, so the above formula still
requires that the $k$-mesh spacing be fine enough that each $|\phi^{p}|<\pi$
for all $k_{p}$, just as was required for $\Delta A_{i}$.

\subsection{\label{subsec:Berryfluxdiag}Berry flux diagonalization}

With Eqs.~(\ref{eqn:DP}), (\ref{equ:pjk}) and (\ref{equ:Phi-phi}), one can
compute the polarization difference using arbitrarily chosen gauges for systems
$A$ and $B$. However, there is still a requirement that the $k$-mesh be fine
enough that all $\phi_{p}$ in Eq.~(\ref{equ:Phi-phi}) are smaller in magnitude
than $\pi$. For a single-band system, this typically does not require a mesh any
finer than that needed to compute $\phi_{\lambda}$ from
Eq.~(\ref{eqn:discrete}). However, the plaquette Berry fluxes $\phi^{p}$ from
Eq.~(\ref{equ:pjk}) are traced over all occupied bands, so their values can
quickly grow much larger in magnitude than $\pi$ when many bands are
contributing.

We can instead decompose each plaquette flux into a sum
$\phi_{p}=\sum_n\phi^{p}_n$ of smaller gauge-invariant phases $\phi^{p}_n$,
where $n$ runs over the number of occupied bands. These are the multi-band Berry
phases or Wilson loop eigenvalues of plaquette $p$, obtained from the unitary
evolution matrix $\mathfrak{U}_{p}$ acquired by traversing the boundary of the
plaquette. Explicitly,
\begin{equation}
\mathfrak{U}_{p} =
{\cal M}^{\braket{12}} {\cal M}^{\braket{23}}
{\cal M}^{\braket{34}} {\cal M}^{\braket{41}}
\label{equ:Udef}
\end{equation}
where ${\cal M}^{\braket{ij}}$ is the unitary approximant of $M^{\braket{ij}}$,
that is, ${\cal M}=V^\dagger W$ where
\begin{equation}
  M=V^\dagger\Sigma W
  \label{eqn:svd}
\end{equation}
is the singular
value decomposition of $M$. The eigenvalues of the unitary matrix
$\mathfrak{U}_{p}$ are of the form $e^{i\phi_{n}^{p}}$, providing the needed
$\phi_{n}^{p}$, which are gauge-invariant.
Since $\mathrm{Im}\mathrm{ln}\det\mathfrak{U}_{p}$ is taken as the Berry flux
  through plaquette $p$, we have in a sense diagonalized this Berry flux
by obtaining the eigenvalues of $\mathfrak{U}_{p}$.
Finally,
the $\phi^{p}_n$ can be summed over all plaquettes to obtain the total
polarization difference via
\begin{equation}
  \Phi = \sum_{p}\sum_{n}\phi_{n}^{p} \,.
  \label{eqn:phi_pt}
\end{equation}
This is our central result.

For the method to
be applicable the two states $\lambda_{A}$ and $\lambda_{B}$ must be similar
enough that the singular values in $\Sigma$ do not become too small (this
corresponds to the continuum-case requirement that the norm of $\gamma^{AB}$ in
Eq.~(\ref{equ:gamma}) should remain nonzero). For agreement with the continuum
case the individual $\phi_{n}^{p}$ must each be much smaller in magnitude than
$\pi$. This condition is typically satisfied with a $k$-mesh density appropriate
for a standard Berry-phase polarization calculation, but the density of the
$k$ mesh could be increased if necessary. These conditions are
further discussed in Section \ref{sec:caveats}.

The above expressions were all written for the one-dimensional case for the sake of
simplicity; the generalization to two and three dimensions is quite
straightforward. Just as is typically done for the computation of the
Berry-phase polarization,
the computation is carried out separately for each string of $k$-points
in the direction of the desired polarization component, and the results are
then averaged over the complementary directions.

Note that while the computation of overlap matrices between neighboring
$k$-points is quite routine, this procedure also requires overlaps between
wavefunctions of corresponding $k$-points at different $\lambda$ values
(typically different structures). The implementation details for this procedure
are discussed in Sec.~\ref{sec:methodology}.

\subsection{Ionic contribution and alignment}

Up to this point, we have focused only on computing the electronic contribution
to the change in polarization for already fixed choices of unit cells at each
$\lambda$. Differences in origin choice and cell orientation between
$\lambda_{A}$ and $\lambda_{B}$ in general will change the decomposition of the
Berry phase polarization into electronic and ionic contributions, and in
particular can alter the Bloch function overlaps in Eq.~(\ref{eqn:overlap}).
\footnote{Small rotations in going from $A$ to $B$
  present no difficulty, since in practice the calculations are done in internal
  (i.e., lattice-vector) coordinates. For the same reason, a change in strain
  state presents no difficulties in principle, but can require some attention to the
  details of indexing of reciprocal lattice vectors.}
The Berry flux diagonalization method is
most robust when structures are aligned to maximize overlaps, and thus keep
elements of the $\Sigma$ matrix in Eq.~(\ref{eqn:svd}) (the singular values)
from becoming too small. We make this choice of unit cell by first aligning the
structures to minimize the root mean squared displacements of the
ionic coordinates. After this initial
alignment, we further refine the choice of origin by translating along the
polarization direction
to maximize the
smallest of all the singular values encountered
while scanning over all $k$-points in the above-described procedure.
This additional refinement can be performed without additional first-principles
calculations using the existing wavefunctions; in the plane-wave
representation this is accomplished by computing
\begin{equation*}
M_{mn}^{(AB)}({\bf k})=\langle\psi_{m{\bf k}}^A\vert T_\mathbf{\tau}\,\psi_{n{\bf k}}^B\rangle
=\sum_{\bf G} C_{m,{\bf G+k}}^{(A)*}
              C_{n,{\bf G+k}}^{(B)} e^{-i{\bf G}\cdot\mathbf{\tau}}
\end{equation*}
where $T_\mathbf{\tau}$ is the extra translation by $\mathbf{\tau}$
and the $C_{n,{\bf G+k}}$ are the plane wave coefficients.

The ionic contribution to the polarization change
is given by
\begin{equation}
  \Delta \vec{P}_{\mathrm{ion}} = \frac{e}{V_{\mathrm{cell}}}\sum_{i}Z_{i}\Delta\vec{r}_{i}
\end{equation}
where $\Delta r_{i}$ is the displacement of ion $i$ between states $\lambda_{A}$
and $\lambda_{B}$.

\section{\label{sec:methodology}Methods}

The Berry flux diagonalization method is a post-processing step for
wavefunctions generated by first-principles density-functional-theory codes. Our
current implementation of the method\cite{git_bflux}
 is for wavefunctions in a plane-wave basis. Here we perform
calculations in ABINIT using the norm
conserving scalar relativistic ONCVPSP v0.3 pseudopotentials with the LDA
exchange correlation functional \cite{PhysRevB.88.085117}. The necessary
overlap matrices are computed from the NetCDF wavefunction files produced by
ABINIT, read using the abipy library \cite{GONZE2019107042} (https://github.com/abinit/abipy).
The pymatgen library\cite{ONG2013314} is used in the process of computing the ionic contribution.

We validate and demonstrate the Berry flux diagonalization method as follows.
First, we use the method to compute the switching polarization of the
prototypical ferroelectric perovskite oxides BaTiO$_{3}$, KNbO$_{3}$, and
PbTiO$_{3}$, for which the computation of the switching polarization by existing
methods is straightforward. We then compute the switching polarization of pure
PbTiO$_{3}$ and PbTi$_{0.75}$Zr$_{0.25}$O$_{3}$ with a $2\times2\times1$
supercell, using both the present method and the fictitious minimal path method
for direct comparison. The atomic positions in PbTi$_{0.75}$Zr$_{0.25}$O$_{3}$
were taken to be the same as in the pure system.

\section{\label{sec:results} Results}

The computed switching polarizations for the prototypical ferroelectric
perovskite oxides BaTiO$_{3}$, KNbO$_{3}$, and PbTiO$_{3}$, are 0.26 C/m$^2$,
0.29 C/m$^{2}$, and 0.77 C/m$^{2}$ respectively, resolving the branch choice in
the Berry phase differences with the quantum of polarization for the primitive
cells being 1.04 C/m$^2$, 1.03 C/m$^2$, and 1.08 C/m$^2$, respectively. These
switching polarization values are in complete agreement with reported
first-principles values obtained by the fictitious minimal path approach, which
have previously been established to be in good agreement with experimental
observations \cite{PhysRevB.96.035143}. In this section, we give a detailed
analysis of the results for pure PbTiO$_{3}$, which has the largest polarization
and thus presents the most difficulties of the three. We do this for three
cases, namely in the primitive 5-atom cell, in a $2\times2\times1$ supercell,
and in the same supercell, but with one Ti replaced by Zr.

\begin{figure}
  \includegraphics[width=\linewidth]{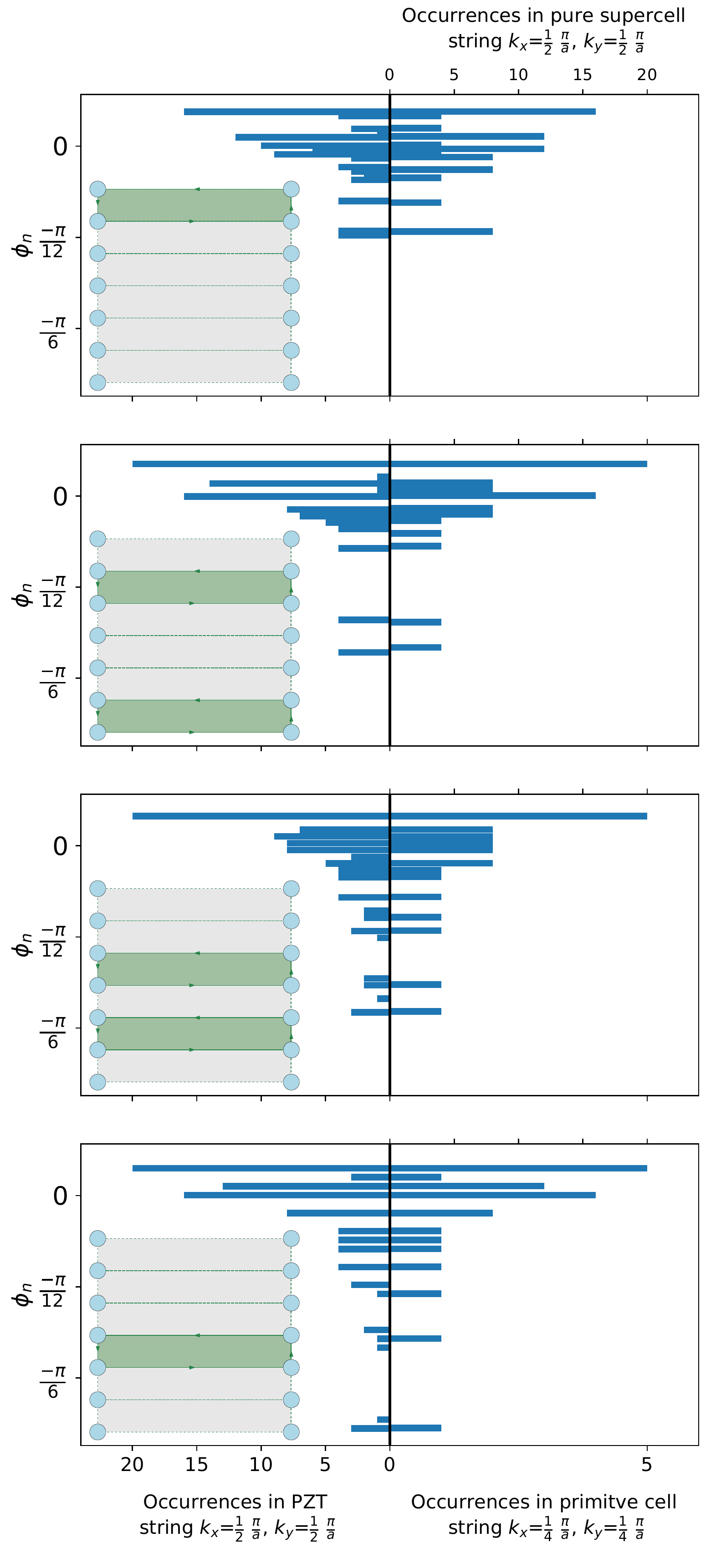}
  \caption{Histogram of Wilson loop eigenvalues ($\phi_{n}^{p}$ from
    Eq.~(\ref{eqn:phi_pt})) for the plaquettes highlighted in the insets following
    the form of Fig.~\ref{fig:k_lambda_space}. In each of the two middle plots,
    the two highlighted plaquettes have identical contributions due to time
    reversal symmetry. Values for pure PbTiO$_{3}$ are shown on the right. The
    occurrences of values for the primitive and supercell systems differ only by
    a factor of 4 as indicated by the two axis scales at the top and bottom of
    the figure. Values for PbZr$_{0.25}$Ti$_{0.75}$O$_{3}$ are shown at
    left.}
  \label{fig:wlevs}
\end{figure}

\begin{figure}
\includegraphics[width=\linewidth]{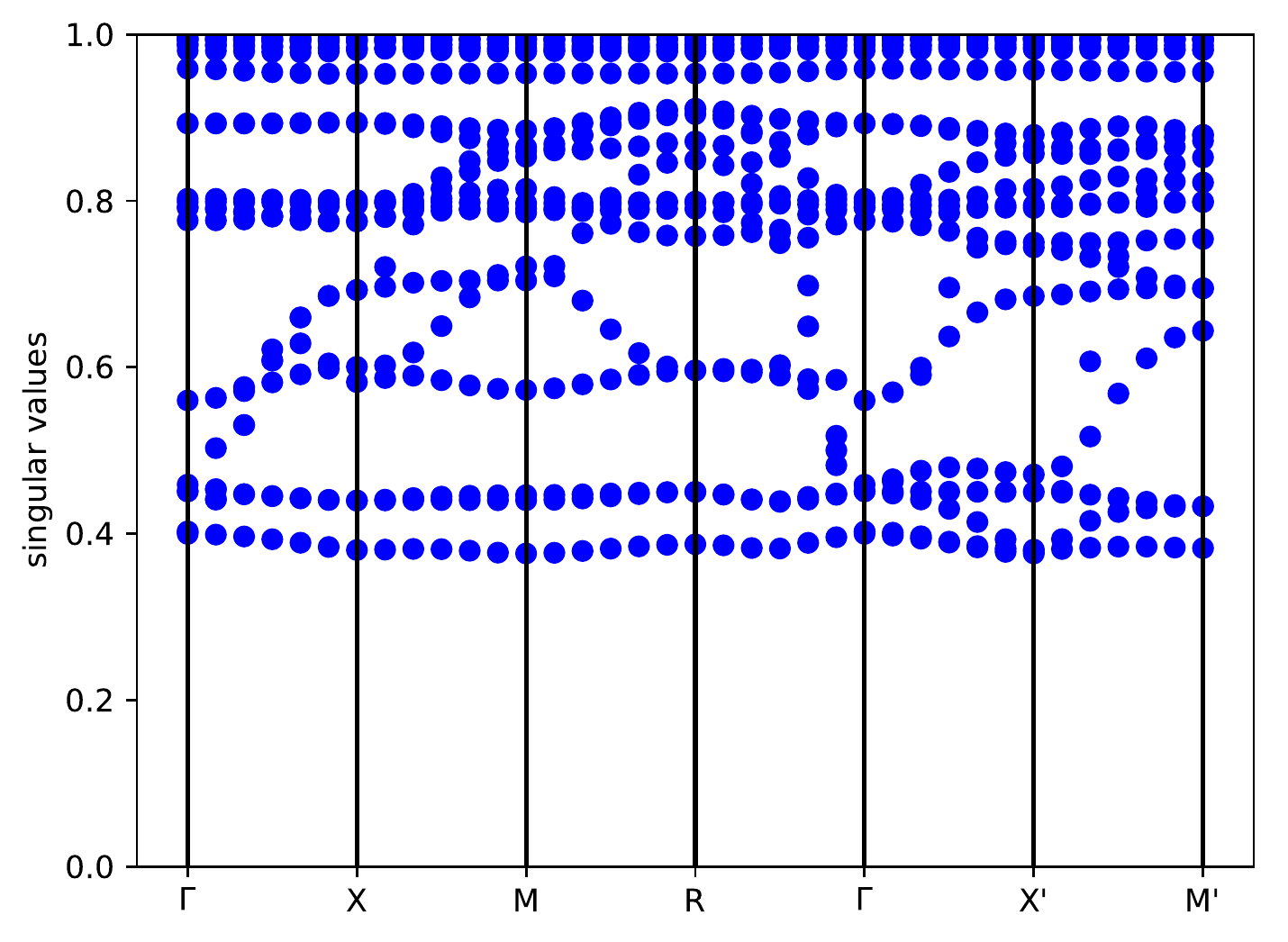}
\caption{Singular values throughout the Brillouin zone for PbTiO$_{3}$, sampled on a 12x12x12 $\Gamma$ centered $k$ mesh.}
\label{fig:sing_vals}
\end{figure}

The key quantities here are the Wilson loop eigenvalues, which are summed in
Eq.~(\ref{eqn:phi_pt}) to obtain the change in polarization. For PbTiO$_3$, the
distribution of the Wilson loop eigenvalues is shown in Fig.~\ref{fig:wlevs} for
plaquettes along the string of $k$-points corresponding to
$k_{x}= \pi/4a, k_{y}=\pi/4a$ for the primitive cells, and to the corresponding
point $k_{x}= \pi/2a, k_{y}=\pi/2a$ for the supercell systems. All Wilson loop
eigenvalues are found to be much smaller in magnitude than $\pi$, mostly
clustered around zero, with a bias in the direction of the electronic
polarization change. Here this is negative given the choice of initial and final
states.

Each individual contribution to the change in polarization for the supercell is
identical to that of the primitive cell, except that they appear with
multiplicity four due to the translational symmetries
that were lost in the
supercell system. So, while the change in dipole moment for the supercell is
four times as large as that for the primitive unit cell, and is thus
significantly larger than the $2\pi$ phase ambiguity, this does not present any
difficulties in the Berry flux method.

The Wilson loop eigenvalues for the system with one Ti replaced by Zr is shown
in the left portion of Fig.~\ref{fig:wlevs}. All eigenvalues fall in the same
range as the pure PbTiO$_{3}$ system,
but with some splitting of values. The
switching polarization for the system with Zr was found to be 0.762 C/m$^{2}$
compared to the slightly larger 0.771 C/m$^{2}$ of the pure system.

In Fig.~\ref{fig:sing_vals}, we show the singular values of overlap matrices $M$
between initial and final states at corresponding $k$-points for PbTiO$_{3}$ in
its primitive cell. These singular values are the diagonal elements of $\Sigma$
from Eq.~(\ref{eqn:svd}).
If the singular values do
not approach zero at any point in the Brillouin zone,
the computed information for initial and final states
determines the polarization change within the
same gauge class assumption.
Fig.~\ref{fig:sing_vals} shows that the singular values for PbTiO$_{3}$ are well
behaved.

\section{\label{sec:discussion} Discussion}
\subsection{\label{subsec:compare} Comparison to fictitious path approach}

In this section we compare the Berry flux diagonalization method to the commonly
used fictitious path approach \cite{PhysRevB.71.014113,
  uratani05_first_princ_predic_giant_elect_polar,
  fu13_diisop_bromid_is_high_temper}, using PbTiO$_{3}$ in its primitive cell
and in a $2\times2\times1$ supercell as illustration.

\begin{figure}
  \includegraphics[width=\linewidth]{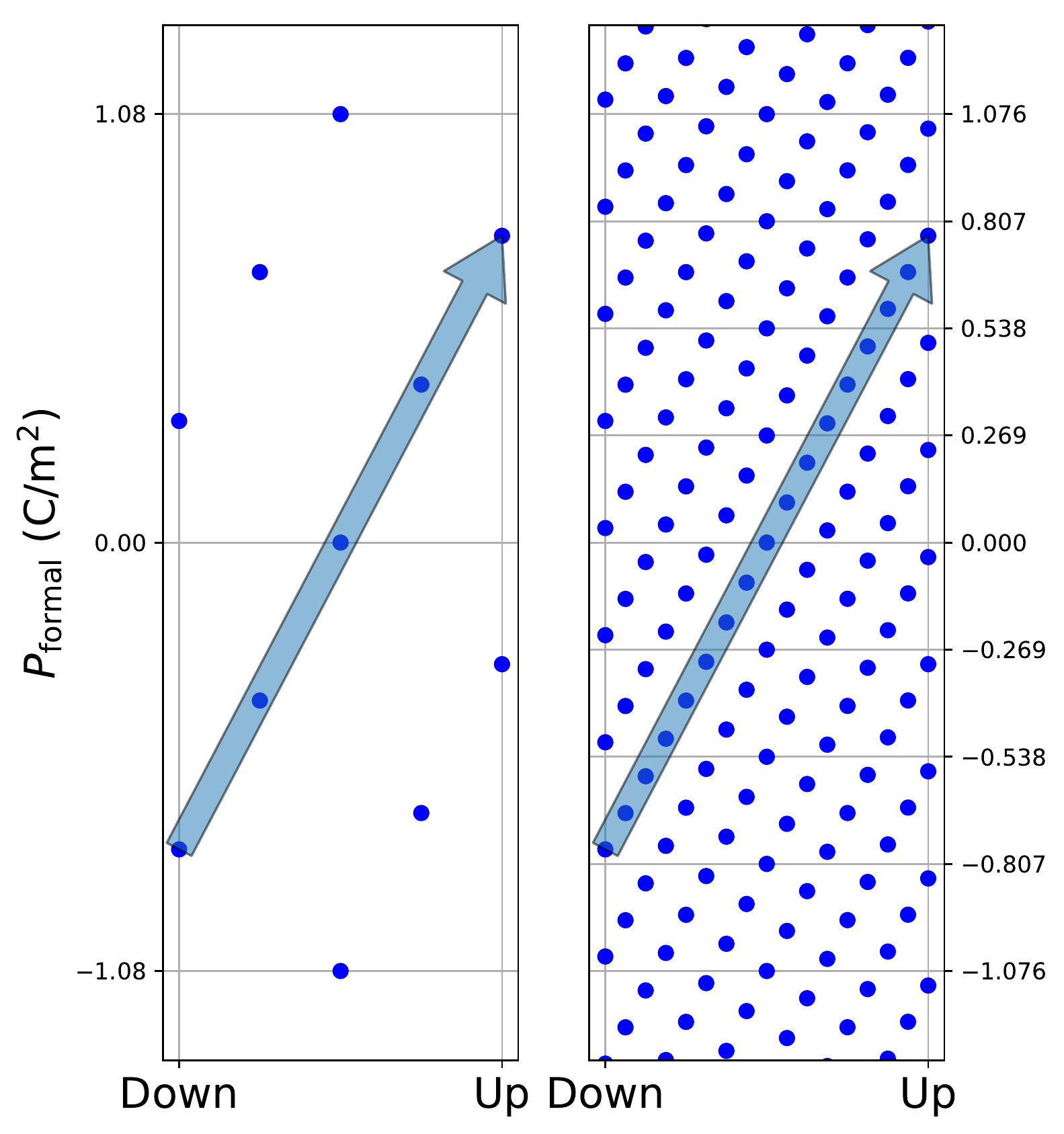}
  \caption{Evolution of formal polarization of PbTiO$_{3}$ along a linearly
    interpolated switching path for the primitive cell (left) and a
    $2\times2\times1$ supercell (right). Ticks and horizontal lines mark the
    polarization quantum. The blue arrow indicates the change in polarization,
    which with the Berry flux diagonalization method only requires calculations
    in the initial state and symmetry-related final state.}
  \label{fig:old_method}
\end{figure}

For the fictitious path approach, we choose a simple linearly interpolated path
between oppositely polarized states.
Fig.~\ref{fig:old_method} shows the formal polarization which is determined
modulo the polarization quantum, computed at points along the path for two
different sampling densities. Starting with an arbitrary choice for the initial
state, the branch is chosen by connecting to the closest value for the next
sampled state along the path. The difference between the final and initial
states is then divided by two to get the spontaneous polarization.

For the case of the primitive cell, calculations for three intermediate states on
the path are needed correctly to resolve the branch ambiguity. In the case of
the supercell, because of the four-fold decrease in the polarization quantum, the
number is significantly larger: 15 intermediate calculations must be done to
resolve the branch ambiguity. The Berry flux diagonalization approach in both
cases, shown by the blue arrow, predicts the change in polarization (with the
correct branch choice) using only the wavefunctions in the initial and final
states.

We note that other approaches have been discussed that utilize partial
information in addition the evolution of $P_{\mathrm{formal}}$, such as nominal
valence charges and Born effective charges \cite{vanderbilt_2018_3}.
This additional information can help determine the choice of polarization value
at the next point on the path even when this is not the smallest change,
reducing the sampling density needed. However, the implementation tends to be
ad-hoc and is not suitable for automated high-throughput applications. Furthermore,
such approaches may not be reliable in situations where these assumed charges are
not constant through the switching process.

\subsection{\label{sec:relation} Relation to Wannier functions}

The Wilson loop eigenvalues $\phi_{n}^{p}$ used in Eq.~(\ref{eqn:phi_pt}) and
shown in Fig.~\ref{fig:wlevs} have a close relation to the position expectation
value of maximally localized Wannier functions, which we refer to maximally
localized Wannier centers\cite{BLOUNT1962305}. The parallel transport procedure
used in obtaining these $\phi_{n}^{p}$ is precisely the same as that used in
obtaining the Wannier centers maximally localized along one dimension. In the
Berry flux diagonalization method the procedure is performed around the
plaquettes, while when computing maximally localized Wannier functions the
procedure is performed across the loop formed by traversing the Brillouin zone
at a single $\lambda$. In the Wannier case, the Wilson loop eigenvalues obtained
are complex numbers with phase $2\pi r_{n}/a$, with the $r_n$ being the
maximally localized Wannier centers. \cite{PhysRevB.56.12847}. The $r_n$ can be
treated as the positions of point charges to compute the formal
polarization\cite{PhysRevB.56.12847}. Similarly, the Wilson loop eigenvalues
obtained in the Berry flux diagonalization method can be understood as
contributions to changes in positions of point charges, yielding the change in
formal polarization.

\subsection{\label{sec:caveats} Conditions for applicability}

To make the correct branch choice and compute the change in polarization, some
assumption about the dynamics of the switching process must be made. In the
method presented in this work, the assumption is that the system evolves in some
minimal way between oppositely polarized states, based on the empirical fact
that computation of the polarization change along a fictitious minimal path
generally corresponds to the measured value. Ionic contributions to the change
in polarization are separated by assuming displacements are minimized, and
electronic contributions are separated by assuming that as the wavefunctions
evolve along the physical switching path, they maintain a high degree of
overlap.

This regime where the latter assumption breaks down can be detected automatically. When
the changes in the electronic states across changes in $\lambda$ becomes large, the
overlaps in wavefunctions become small, and some singular values of the $\Sigma$
matrix of Eq.~(\ref{eqn:svd}) approach zero. The implementation of the method
checks to make sure that no singular values
anywhere in the
Brillouin zone fall below a threshold (see Fig.~\ref{fig:sing_vals}).
Numerical experiments have shown that a threshold of around ~0.15 seems to work well
for systems tested. There is of course also a branch ambiguity if the Wilson
loop eigenvalues ($\phi_{n}^{p}$ of Eq.~(\ref{eqn:phi_pt})) have
magnitudes close to $\pi$. In practice, we have found no cases where this happens
without the requirement on the singular values failing first. This can be
understood from the viewpoint that the Wilson loop eigenvalues are related to
displacements of Wannier centers,
with a value
of $\pi$ corresponding to a single charge moving by half a unit cell. When the charge
is moved over such a distance the overlaps tend to become small, especially in
  an insulating system where states are localized.
For such systems, one can revert to constructing intermediate states along
$\lambda$. If each change in polarization is computed using the Berry flux
diagonalization method, $\lambda$ can be sampled more coarsely than methods that
track only the total phase.
However, in doing so one should beware of making possibly unsafe
assumptions about the dynamics of the switching process.

\section{Conclusion}
The Berry flux diagonalization method presented here provides a way to compute
the change in polarization that is more easily automated,
as well less computationally expensive, than existing approaches. The magnitudes
of the singular values obtained in the course of the calculation provide a
built-in test of whether the two systems being compared
are sufficiently similar that
a class of minimal paths producing the same change in polarization can be inferred.
Future work will explore the application of this
method to the change in polarization between two states that are not symmetry
related, such as in pyroelectrics, antiferroelectrics, heterostructures and
insulators in finite electric fields.
It will also be interesting to test the applicability
of the approach to different classes of ferroelectrics, such as organic,
inorganic order-disorder, charge-ordered, or improper ferroelectrics.
Generalizations of the method to
the computation of other quantities requiring Berry curvature integration, such
as Chern numbers and characterization of Weyl points, should also reward future
investigation.

\section{Acknowledgments}
This work was supported by ONR Grant N00014-16-1-2951, ONR N00014-17-1-2770, and
NSF DMR-1334428.
We would like to thank Don Hamann and Cyrus Dreyer for useful discussions.


\bibliography{branch.bib}{}
\end{document}